# Determinants of Self-Interstitial Energetics in Refractory High-Entropy Alloys

Zichen Zhang[1], Zhiling Luo[1], Wang Gao[1]*, Qing Jiang[1]*

[1] Key Laboratory of Automobile Materials (Jilin University), Ministry of Education, Department of Materials Science and Engineering, Jilin University, Changchun, 130022, China.

*Corresponding author(s). E-mail(s): wgao@jlu.edu.cn; jiangq@jlu.edu.cn.

**Abstract**

Self-interstitials play a central role in governing the mechanical and anti-irradiation properties of refractory high-entropy alloys (RHEAs), however, the prediction of interstitial formation energies ($E_f$) is formidable due to the chemically complex environments in RHEAs. Herein, we develop a framework based on the tight-binding model to quantify the effects of complex alloying and lattice distortion on $E_f$. Our scheme reveals that $E_f$ is jointly determined by the average d-band center of RHEAs and the d-band width of interstitial sites. Notably, the d-band width mainly depends on the interatomic hopping matrix and atomic size-determined coordination number, which together make the metallic bonding around interstitials in RHEAs resemble the distance-dependence law of van der Waals forces. By capturing d-band coupling character, our descriptor describes both interstitial configurations within a universal framework. Our model reveals a new physical picture of interstitial formation, providing a useful tool for the design of high-performance RHEAs.

**Teaser**

An analytical model for interstitial formation energies in refractory high-entropy alloy

**Keywords**

Interstitial defects, Formation energy, Analytic model, First-principles calculations

**Introduction**

Body-centered cubic (BCC) High-entropy alloys (HEAs) exhibit exceptional mechanical properties and superior radiation tolerance, (*1–8*), showing significantly reduced defect-cluster densities and void swelling rates under high-dose irradiation because of their inherently loose-packed crystal structures and severe lattice-distortion effects(*9*). Generally, both vacancy and interstitial defects are simultaneously generated in RHEAs, and their formation and evolution directly govern the microstructural evolution and macroscopic property degradation of RHEAs. However, the previous studies have predominantly focused on vacancies in RHEAs(*10*), leaving the interstitial behavior incompletely understood. Particularly, the site-to-site chemical disorder in RHEAs induces complex alloying effects and local lattice distortions around interstitial sites, presenting substantial challenges for understanding interstitial energetics.

Several pioneering studies have investigated the preferential formation ability of different interstitials in RHEAs and attempted to explain the findings with the atomic radii and interstitial electronic structures (*11–13*). In VTaCrW, the V- and Cr-containing-dumbbell interstitials (with the <110> orientation) are dominant (*14*). In AlNbTiZr, the preferential formation of Al-Al dumbbells and the suppressed formation of Zr-Zr dumbbells are observed and rationalized using the charge redistribution induced by the high electronegativity of Al atoms (*15*). However, the previous studies have focused on the statistical analyses of interstitials, while the decisive factors governing the formation energies of interstitials remain elusive. In particular, there still lacks a quantitative analytical model for determining interstitial formation energies, failing to establish a comprehensive physical picture.

In this work, we propose a descriptor $D_{int}$ for the $E_f$ based on the tight-binding (TB) model. $D_{int}$ captures the d-d hoping matrix between the interstitial atoms and their neighboring atoms and the atomic size-determined coordination number. $D_{int}$ essentially reflects the d-band width at the interstitial site. The slope of the linear relationship between the descriptor and $E_f$ is determined by the average d-band center of RHEAs. Our analytical framework reveals an interesting physical picture: interstitial formation energy is governed by the coupling between the local alloying effect at an interstitial site and the global average alloying effect of the alloy. Our model shows broad applicability across different RHEAs and offers a useful tool for designing high-performance RHEAs

**Results**

We systematically investigate the factors governing the formation energies of self-interstitials in RHEAs, including NbMoTaV, NbMoTaTi, NbMoTaCr, NbMoTaHf, NbMoTaRe, NbMoTaW, NbMoTaZr, WTaV, WVCr and HfNbTiV. We find that the stable interstitial configurations can be classified into two types: dumbbell-type interstitials and crowdion-type interstitials, as shown in Fig. 1(a-b). The dumbbell configuration is formed by two atoms co-occupying a single lattice site along the <110> direction, while the crowdion configuration involves an extended distortion of crystal lattice along the <111> direction. The compressive strain in the crowdion core extends its influence over a region encompassing the central 3–4 atoms of the crowdion chain(*16*), we therefore define the crowdion core as the 4 most displaced atoms along the <111> direction. Notably, a single dumbbell atom possesses nine first-nearest neighbors and three second-nearest neighbors (fig. S2a), exhibiting an oversaturated nature compared to bulk atoms. In contrast, a single crowdion atom has eight first-nearest neighbors and six second-nearest neighbors (fig. S2b). Although its coordination number matches that of a bulk atom, the introduction of the extra atom still results in oversaturated coordination for its neighboring atoms. Therefore, the crowdion core region is also a bond-oversaturated zone. Our results show that the crowdion configuration occurs more frequently than the dumbbell configuration in NbMoTaTi, NbMoTaV, NbMoTaZr, and WTaV, whereas the dumbbell structure dominates in the other studied systems (Fig. 1c).

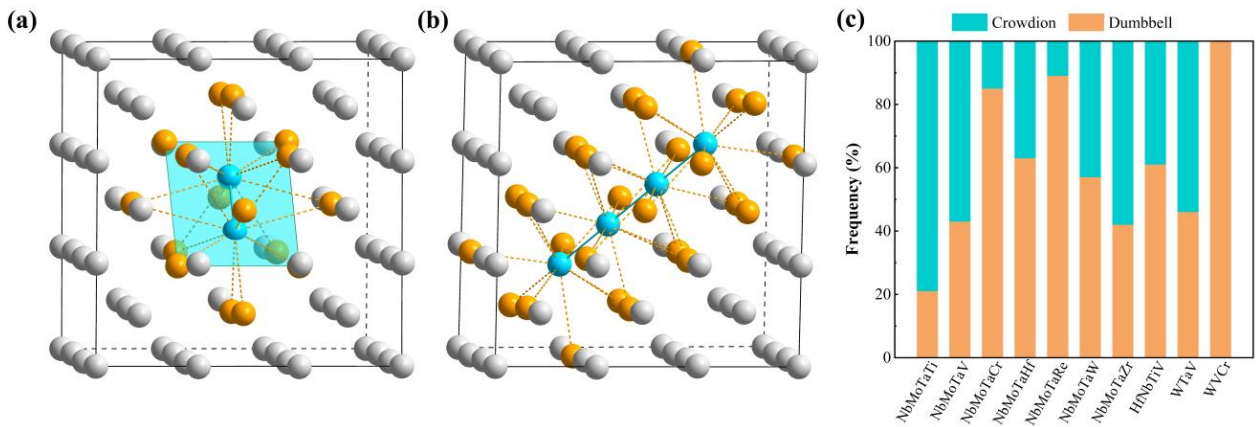

**Fig.1 The two configurations of stable interstitial structures in RHEAs.**
The blue spheres represent atoms located at the distortion center, while the orange spheres denote their paired neighboring atoms. (a) represents a two-atom dumbbell structure oriented along the <110> direction. (b) represents a <111>-oriented crowdion structure, in which the distortion center involves four atoms. (c) shows the fraction of dumbbells and crowdions in each system under study, with blue bars indicating crowdions and orange bars indicating dumbbells.

**The interactions between the interstitial atoms and its neighboring atoms**

The stability of interstitials in RHEAs is governed by the interactions between the interstitial central atoms and their neighboring atoms. We will draw inspiration from the spirit of the TB model to understand these interactions. The TB model shows that in TMs the s-bands are delocalized and half-filled while the d-bands are localized, thereby the d-band coupling dominates the interatomic bonding trend with the s-band contribution as a constant, from one metal to another. We first assume that this rule approximately works for the interstitials of RHEAs. Thus, the interactions between the interstitial atoms and their neighboring atoms mainly arise from the corresponding d-d coupling, for which the d-band width $W_d$ plays a dominant role according to the TB model(*17*, *18*). From the moments theorem, $W_d$ is proportional to the coordination number (*Z*) and hopping integral (*V*) in TMs as $W_d \propto Z^{\frac{1}{2}}|V|$, in which *Z* is generally taken as an integer although the change of *Z* is accompanied with the change of bond length(*17*). Indeed, a neighboring atom with the longer interatomic distance should be count as less than a bulk neighbor, generating a continuous function of *Z* as a function of the interatomic distance. Namely, the change of bond length can significantly alter the bonding strength as well as *Z*, which is particularly important for self-interstitials in RHEAs: the interstitials experience severe lattice distortion and usually have the local oversaturated bonds and the enlarged interatomic distance (by > 12%) compared to the pristine RHEAs. $Z^{\frac{1}{2}}$ was found to depend on the bond order of metal atoms (*s*) in $Z^{\frac{1}{2}} \propto s$, reflecting the behavior of the square-root bond-cutting model for the bonding strength of surface atoms. Since *s* corresponds to the bonding strength and depends on the interatomic distance following $s=1/\exp((R-R_0)/0.37)$ in the framework of Pauling's valence bond theory(*19*), it can be taken as $s \propto \frac{1}{d}$ with the first-order approximation. Combining with the hopping integral *V* from Muffin-Tin-Orbital (MTO) theory, we obtain the descriptor $D_{Ii}$ quantifying the interactions between an interstitial atom and one of its neighboring atoms in RHEAs,

$$D_{Ii} = |\eta_{ddm}| \frac{(r_d^I \times r_d^i)^{\frac{3}{2}}}{d_{Ii}^6} \quad (1)$$

where, $r_d$ denotes the d-band radius of TMs, characterizing the spatial extent of the d-orbitals. The value of $r_d$ for each TM element is obtained from solid-state table in literature(*20*). $d_{Ii}$ represents the internuclear distance between the interstitial atom and a given neighboring atom. The coefficient $\eta_{ddm}$ is governed by to the type of d-d orbital bonding interaction, with the σ-bond coefficient ($|\eta_{dd\sigma}| = 16.2$) being approximately twice that of the π-bond ($|\eta_{dd\pi}| = 8.75$).

**Construction of Electronic Descriptors for Interstitial Formation Energy**

The complex alloying effects around interstitials are characterized by the descriptor $D_{int}$, defined as the summation of $D_{Ii}$ between the interstitial atoms and their neighboring atoms,

$$D_{int} = \sum_{i=1}^{n_1}\sum_{j=1}^{n_2} D_{ij} - \sum_{k=2}^{n_1} D_{k-1,k} \quad (2)$$

$n_1$ and $n_2$ denote the number of interstitial atoms and the number of neighboring atoms for each interstitial atom. The first term represents the summation of $D_{Ii}$ between the interstitial atoms and all its neighboring atoms, while the latter denotes that between the interstitial atoms.

Our descriptor exhibits linear correlations with the $E_f$ of dumbbell and crowdion configurations in in all considered RHEAs (Fig. 2a-c and fig. S1), with $E_f = kD_{int} - b$ (*k* is the slope and *b* is the intercept). It is evident that, when the possible difference in d–d coupling forms between dumbbells and crowdions, namely the contribution of $\eta_{ddm}$, is neglected, the dumbbell and crowdion comply with two distinct linear trends, with the slope *k* of dumbbells systematically larger than that of crowdions. This indicates that the difference in d-orbital coupling induced by the interstitial geometry significantly affect $E_f$.

We then study the difference of d–d coupling between dumbbells and crowdions. In dumbbell cores, the extremely short interatomic distance between the two core atoms localizes the bonding electrons and promotes a strongly directional head-on overlap of d-orbital, corresponding to σ-type bonding. In the case of crowdions, the lattice strain induced by the inserted atom is shared by four atoms along the <111> atomic string, leading to a significant lattice distortion and a more delocalized distribution of bonding electrons. As a result, the d-d overlap in crowdions exhibits a predominantly π-type bonding character, because of the more tolerant to lattice distortion nature of π-like overlap. We therefore use the coefficients $|\eta_{dd\sigma}| = 16.2$ for dumbbells and $|\eta_{dd\pi}| = 8.75$ for crowdions according to the TB model.

As shown in Fig. 2d-f, once the σ/π bonding character of interstitial cores is taken into account, dumbbells and crowdions coalesce to follow a single linear relation. These results reveal an important electronic-structure contrast between the two interstitial configurations in RHEAs: the d-d coupling is dominated by σ-type bonding in dumbbells but by π-type bonding in crowdions. Accounting for this difference allows the two defects, even with different numbers of atoms in their interstitial cores, to be described within a unified linear framework.

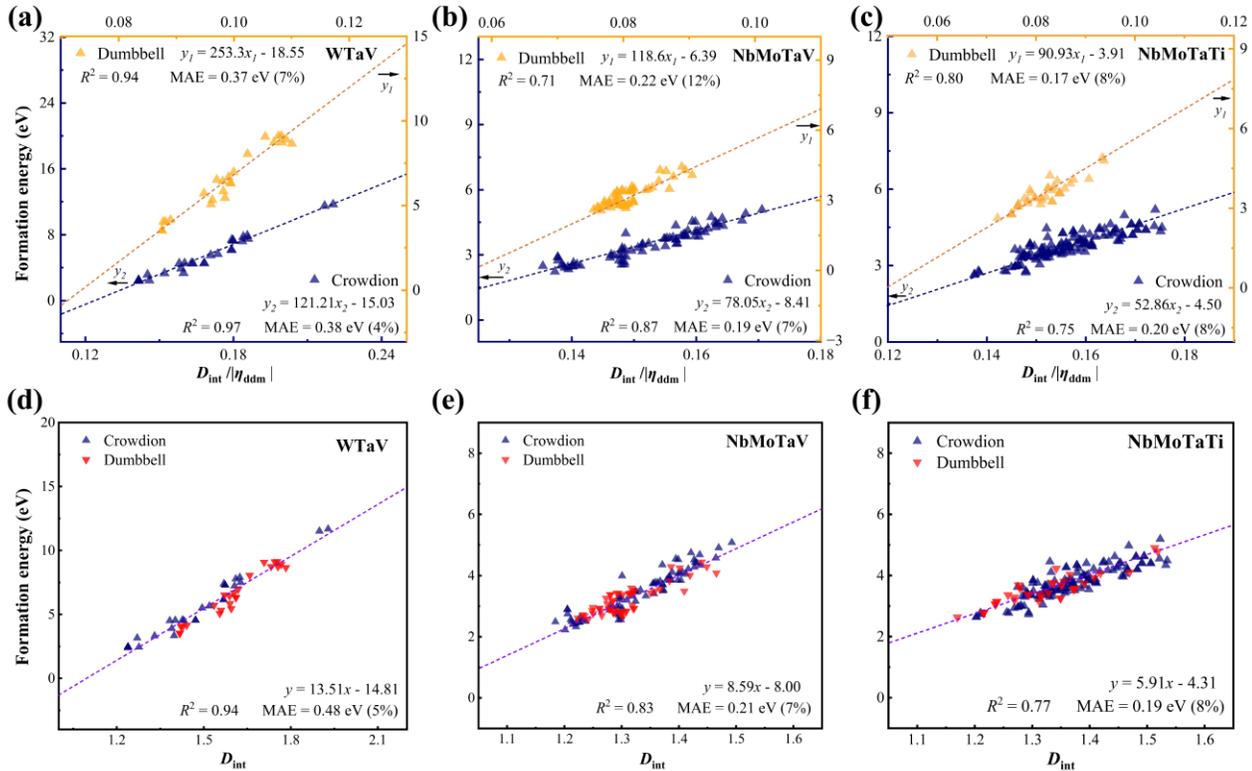

**Fig.2 The relationship between the descriptor $D_{int}$ and the interstitial formation energy.**
(a-c) The linear relationships between the descriptor (without considering the d–d coupling coefficient) and the formation energy in WTaV, NbMoTaV, and NbMoTaTi. The orange symbols denote dumbbell configurations, whereas the blue symbols denote crowdion configurations. (d-f) Unified linear relationship of $E_f$ for dumbbells and crowdions using the descriptor containing d–d coupling coefficient

We now turn to examine the origin of the slope and intercept in the linear scaling relation determined by $D_{int}$. As shown in Fig. 3a, the slope $k$ exhibits a well-defined linear correlation with the intercept $b$, indicating that the slope and intercept have the same physical origin. The expression is given as $b = \frac{3}{2}k - 5$. Clearly, they are independent on $D_{int}$, namely the local alloying effect characterized by the d-band width, meanwhile they are unchanged for any interstitial site. We thus infer that k originates from the intrinsic average alloying effect for a given RHEA, which likely rely on other d-band properties, such as the d-band center and d-band occupation, rather than d-band width. Consequently, we employ the average value $\bar{\psi}$ of the descriptor $\psi = \frac{S_v^2}{\chi}$, which reflects the d-band center from our previous work (21) to quantify this global alloying effect. $S_v$ and $\chi$ denote the valence electron number and electronegativity, respectively. We find that $\bar{\psi}$ reasonably captures the trend of the slope k across different RHEA systems (Fig. 3b),

$$\bar{\psi} = \frac{1}{n} \sum_{i=1}^{n} \psi_i \quad (3)$$

Here, $\psi_i$ denotes the $\psi$ of the i-th constituent element in a RHEA, and $n$ is the total number of components. Recalling the relationship between slope $k$ and intercept $b$, we now obtain the formula of interstitial formation energies of RHEAs,

$$E_f = (1.6\bar{\psi} - 17)(D_{int} - 1.4) + 4 \quad (4)$$

As shown in Fig. 3c, the mean absolute error (MAE) of prediction based on Eq. (4) is 0.56 eV (5%). The average accuracy across all systems reaches an MAE of approximately 0.25 eV if $E_f$ is predicted for individual system directly using the $D_{int}$ descriptor.

These findings reveal an interesting physical picture of self-interstitial formation in RHEAs, where the average alloying effect controls interstitial stability via the position of the d bands. These results demonstrate that the bonding of interstitials in RHEAs share the similarities with the adsorption of molecules on metal surfaces, both depending on the position of d-bands. This behavior likely stems from the supersaturated bonds of interstitials in RHEAs, which makes the interatomic bonding of metal atoms around interstitials (in bulk region) resemble that between molecules and metal surfaces. This similarity (namely the dominant role of d-band center) is only limited to the global alloying effect, while the local alloying effect of interstitials is determined by the d-band width. Overall, these results suggest that the stability of self-interstitials in RHEAs is determined by the interplay between the global average alloying effect of RHEAs and the local alloying and geometric effect at the interstitial sites, namely the average d-band center of RHEAs and the d-band width of interstitial sites.

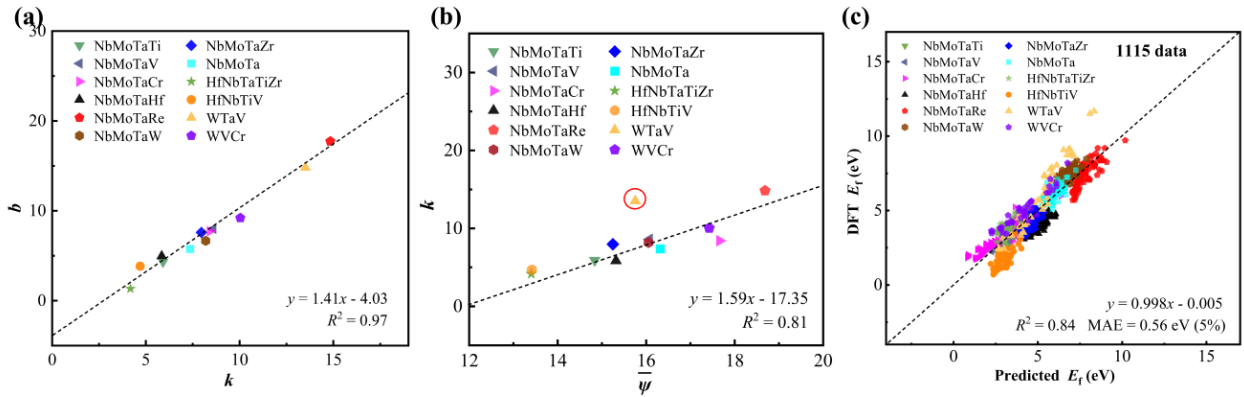

**Fig 3. The physical origin of the slope and intercept in the linear scaling of interstitial formation energies**

(a) The correlation between $k$ and $b$ across different RHEA systems. (b) The relationship between the descriptor $\bar{\psi}$ and $k$. The fitted result is obtained after removing the clear outlier (WTaV). (c) Relationship between the predicted values based on our model and DFT calculated values for all RHEAs we investigate. The dashed line represents that the predicted values are the same as the DFT calculation ones.

**Discussion**

To further understand the power law of our descriptor $D_{Ii}$ with respect to the interatomic distance, we test the different powers of distance. Clearly, the power of six exhibits the optimal accuracy in predicting the formation energy of intersitials (see Fig. 4), reflecting the crucial role of the lattice distortion in determining the coordination number and bond properties of interstitials. This behavior reveals a novel physical picture for the interatomic bonding of interstitials in RHEAs. It is known that in the TB model, the d-d interactions in TMs exhibit a distance dependence of $1/L^5$, which is close in its decay rate to but different obviously in law from the $1/L^6$ dependence of van der Waals (vdW) forces. In comparison, for the interstitials in RHEAs that experience oversaturated bonds and severe lattice distortion, the metallic bonding of interstitials displays a distance dependence identical to that of vdW forces, although the strength coefficients of the two exhibit a drastic difference, corresponding to the significantly different bonding strength of metallic and vdW bonding. The metallic bonding of interstitials is thus more localized compared with that of pristine RHEAs, indicating that the interstitial' influence is highly localized in RHEAs.

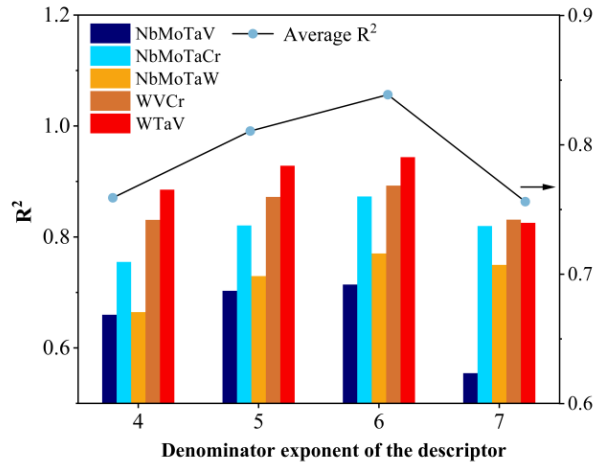

**Fig 4. Relationship Between Denominator Power of Descriptor $D_{Ii}$ and $R^2$ across Systems**
Relationship between the denominator exponent of the descriptor $D_{Ii}$ and prediction accuracy in five alloys (NbMoTaV, NbMoTaCr, NbMoTaW, WVCr, and WTaV).

The descriptor $D_{Ii}$ enables us to evaluate the impact of specific neighboring elements on the formation energy of a given interstitial site. The descriptor $D_{Ii}$ characterizes the properties of an individual atom pair. By isolating the contribution of each atom, the influence of a specific element at a given site can be described by the single-atom descriptor $D_i = \frac{r_d^{1.5}}{R^3}$. For the interstitials in RHEAs, an element with a large $D_i$ increases its $D_{Ii}$ values with neighboring atoms, thereby increasing the formation energy as the slope of $D_{Ii}$ and $E_f$ correlations is positive.

This principle has been validated by numerous studies. The first important example is that in experiments Cr–V–rich particles were observed in the WTaVCr alloy under 1050 K and 3-MeV Cu$^+$ irradiation(22). In contrast, the quinary WTaVCrHf alloy shows a uniform elemental distribution under similar temperature and irradiation conditions, with no secondary phase precipitation(23). This difference can be well understood with our framework. In WTaVCr, the $D_i$ of Cr and V are close to each other ($D_{Cr}$=0.530 and $D_V$=0.529), while W and Ta form another similar group ($D_W$=0.653, $D_{Ta}$=0.640), with the Cr–V values being much lower than those of W and Ta. This produces a highly non-uniform distribution of interstitial formation energies. The elements with low $D_i$ therefore tend to enrich around interstitial sites, lowering the local energy and promoting the nucleation of Cr–V–rich precipitates observed in experiments. Hf exhibits the $D_i$ value ($D_{Hf}$=0.576) that lie between those of V/Cr and W/Ta, which thus can bond with the other four elements and is relatively uniformly distributed in WTaVCrHf. As a result, the introduction of Hf into WTaVCr makes the interstitial configurations that were originally easy to form less favorable, but those that were difficult to form more accessible. In this sense, a uniform distribution of Hf disrupts the strong V–Cr pairing tendency (that inhibits the precipitation Cr–V–rich particles), weakens the short-range order, and effectively smooths the energy landscape. The same mechanism also explains the high irradiation resistance of WTaVCrHf. Hf narrows the distribution of interstitial formation energies, making the distribution of interstitial formation energies much more likely to overlap with that of vacancy formation energies, and thus promotes the recombination between these two types of point defects.

In addition, many theoretical-calculation results are consistent with our model. In V-containing alloys (WV, WTaV, WTaVMo, and WTaVMoNb), the study using molecular dynamics simulations and a machine-learned interatomic potential has shown that nearly all interstitial dumbbells are V–V pairs, and V atoms exhibit a strong tendency to segregate toward interstitial region. (*12*). As indicated by our model, V exhibits the smallest $D_i$ among the constituent elements of these alloys. This suggests that the enrichment of V at the interstitial central sites or their nearest-neighboring shell reduces the interstitial formation energies. As a result, V atoms preferentially pair to form V–V dumbbells and accumulate around interstitial regions. A similar trend is also observed in Cr-containing alloys and NbZrTi alloys(*14*, *24*), as our descriptor suggests.

In summary, we construct a descriptor $D_{int}$ based on the classical tight-binding model to quantify interstitial formation energies ($E_f$) in RHEAs. This descriptor identifies the three primary factors governing the local bonding around interstitials : the d-d hopping matrix between an interstitial atom and its neighbors, the size effect arising from local lattice distortions and the d-orbital overlap mode determined by the interstitial configuration. These three factors together make the metallic bonding around interstitials resemble the distance-dependence law of vdW forces, although the corresponding bonding strength is significantly different. By incorporating the global average alloying effect of systems, we find that $E_f$ is controlled by the coupling between the d-band width of interstitial sites and the average d-band center of RHEAs. Our findings provide an instructive physical picture for understanding the interstitial formation in RHEAs and offer key tools for designing radiation-resistant high-entropy alloys.

**Materials and Methods**

All density functional theory (DFT) calculations in this study were carried out using the Vienna Ab initio Simulation Package (VASP)(*25*), employing the projector-augmented wave (PAW) method (*26*)and the Perdew–Burke–Ernzerhof (PBE) exchange-correlation functional(*27*). We constructed $4 \times 4 \times 4$ BCC supercells containing 128 atoms using the special quasi-random structure (SQS) method(*28*, *29*). Each elemental species in the system was treated as the interstitial atom and inserted into the eight tetrahedral and eight octahedral sites located at the center of the relaxed SQS structure. Structural optimizations were carried out using a plane-wave energy cutoff of 400 eV, and the conjugate gradient algorithm was used with a convergence threshold of 0.02 eV/Å in Hellmann-Feynman force on each atom. The Monkhorst-Pack k-point sampling mesh density was 3 by 3 by 3(*30*).

Interstitial formation energy was calculated by
$$E_f = E_I - E_0 - \mu_i \tag{5}$$
$E_f$ denotes the formation energy of an interstitial in RHEAs, $E_I$ is the total energy of the relaxed interstitial structure, $E_0$ is the energy of the relaxed perfect SQS structure, and $\mu_i$ represents the chemical potential of the interstitial atom, and is calculated according to the Widom-type substitution technique(*15*).

**References**
1. P. J. Barron, A. W. Carruthers, J. W. Fellowes, N. G. Jones, H. Dawson, E. J. Pickering, Towards V-based high-entropy alloys for nuclear fusion applications. *Scripta Materialia* **176**, 12–16 (2020).

**Acknowledgments**
**Funding:** This work was supported by the National Natural Science Foundation of China (Nos. 22173034, 11974128, 52130101), the Opening Project of State Key Laboratory of High-Performance Ceramics and Superfine Microstructure (SKL202206SIC), the Program of Innovative Research Team (in Science and Technology) in University of Jilin Province, the Program for JLU (Jilin University) Science and Technology Innovative Research Team (No. 2017TD-09), the Fundamental Research Funds for the Central Universities, the computing resources of the High-Performance Computing Center of Jilin University, China.
**Author contributions:** Wang Gao and Qing Jiang conceived the original idea and designed the strategy. Wang Gao derived the models with the contribution of Zichen Zhang. Wang Gao and Zichen Zhang analyzed the results with the contribution of Zhiling Luo. Wang Gao and Zichen Zhang wrote the manuscript together. Zhiling Luo and Zichen Zhang performed the calculations with the contribution of Wang Gao. Zichen Zhang drew figures and prepared the Supplementary Materials with the contribution of Wang Gao. Zichen Zhang, Zhiling Luo, Wang Gao and Qing Jiang have discussed and approved the results and conclusions of this article.

**Declaration of Competing Interest:**
The authors declare that they have no known competing financial interests or personal relationships that could have appeared to influence the work reported in this paper.

**Data and materials availability:**
All data are available in the main text or the supplementary materials.

**Supplementary Materials**
Supplementary Fig.S1-S3